\documentclass[twocolumn]{revtex4}

\usepackage{color} 
\usepackage{graphicx} 
\usepackage{amsmath,amssymb} 

\begin{document}

\title{Temperature response of the polarizable SWM4-NDP water model}

\author{Roman Shevchuk} 
\author{Francesco Rao \footnote[1]{Corresponding author;
E-mail: francesco.rao@frias.uni-freiburg.de; Phone: +49 (0)761 203 97336, Fax:
+49 (0)761-203 97451}}
\affiliation{ Freiburg Institute for Advanced Studies, University of Freiburg, Freiburg, Germany.}

\date{\today}

\begin{abstract}

Introduction of polarizability in classical molecular simulations holds the promise to increase accuracy as well as  prediction power to computer modeling. To introduce polarizability in a straightforward way  one strategy is based on Drude particles: dummy atoms whose displacements mimic polarizability. In this work, molecular dynamics simulations of SWM4-NDP, a Drude-based water model, were performed for a wide range of temperatures going from 170~K to 340~K. We found that the density maximum is located far down in the supercooled region at around 200~K, roughly 80~K below the experimental value. Very long relaxation times together with a new increase in the density were found at even lower temperatures. On the other hand, both hydrogen-bond coordination up to the second solvation shell and tetrahedral order resembled very much what was found for TIP4P/2005, a very good performer at the reproduction of the density curve and other properties of bulk water in temperature space. Such a discrepancy between the density curve and the hydrogen bond propensity was not observed in other conventional water models. Our results suggest that while the simplicity of the SWM4 model is appealing, its current parametrization needs improvements in order to correctly reproduce water behavior beyond ambient conditions.

\end{abstract}

\maketitle

\section{Intro}

Molecular simulations provide way to look at water motion and structure at the nanoscale with atomistic details. Notwithstanding classical models demonstrated to correctly reproduce a wide range of observables \cite{Vega2011}, several properties of this fascinating element are not well described yet, including the nucleation mechanism \cite{Matsumoto2002, Moore2011} and the correct proportion between melting and density maximum temperatures to name a few \cite{Vega2005-2}. 

Classical water potentials are attractive because they are cheap to compute as compared to  purely ab-initio approaches. However, to be fast they went through a series of approximations including  rigid molecular structure and on-site fixed partial charges. Like a too short blanket, parametrization of those classical models allowed the correct prediction of some properties leaving behind other ones and vice versa: models like TIP4P-ICE better fit the properties of ice \cite{Abascal2005ice}, others the density anomaly (TIP5P \cite{Mahoney2000}, TIP4P-Ew \cite{Horn2004} and TIP4P/2005 \cite{Abascal2005}) or the diffusion constant  (SPC-E \cite{Mark2001}), but all of them fail to reproduce the broader spectrum of water properties. Although many have agreed that four site potentials might represent the best compromise for  classical water models \cite{Vega2005,Vega2011,Horn2004}, still some important ingredients are missing in these representations: one being \emph{polarization}. In common words, the latter refers to the ability of an atom to change its charge in response to the environment, reflecting a redistribution of the electronic cloud. This effect is pretty obvious and omnipresent when dealing with charged atoms. Think for example of the effect of an ion on the charge distribution of the surrounding molecules \cite{Lamoureux2006_ion}. The drawback for the introduction of polarizability in a classical potential is however two fold. First, given the increased number of degrees of freedom a fully fledged polarizable molecular model is much more computationally expensive to calculate. Second, parametrization of such a model is non-trivial \cite{Brooks2009}.

In recent years, we saw the rise of several polarizable water models such as for example BK \cite{Baranyai2010}, AMOEBA \cite{Ren2003} and  SWM4 \cite{Lamoureux2003}. They differ among each other in the way polarization is implemented. In BK  the charge distribution is represented by Gaussian functions while polarizability  is introduced via a charge-on-spring method \cite{Baranyai2010}. In AMOEBA, polarization effects are treated via mutual induction of dipoles with experimentally derived polarizabilities and a 14-7 potential  to treat Van der Waals interactions \cite{Halgren1992}. A new version of this potential called iAMOEBA (where the ''i'' stands for inexpensive) \cite{Wang2013} makes this model only four times slower compared to a conventional water model. Finally, another way to introduce polarization is to use a Drude oscillator potential \cite{Vanmaaren2001,Lamoureux2003}. In this case a point charge is connected via a classical spring to the oxygen atom via a dummy atom where an external field displacing the dummy particle would in turn induce a dipole \cite{Lamoureux2003}. A fairly adopted implementation of this solution is represented by the SWM4-NDP model \cite{Lamoureux2006}. Being a pairwise based potential and resembling the architecture of a regular four site model like TIP4P, this model seems to fit better into  a conventional molecular force field framework. The model is certainly promising at ambient conditions showing better agreement with experiments for viscosity \cite{Stukan2013} and hydration of the calcium carbonate \cite{Bruneval2007}.

Here, we make an effort to further explore the behavior of the SWM4-NDP model on a wider temperature range. Focusing on some basic properties of bulk water, extensive molecular dynamics simulations were performed for temperatures ranging from 170~K to 340~K. We aimed at the characterization of the density curve as well as at the hydrogen bond propensities and tetrahedral order. The model does not seem to perform very well in terms of density, especially in the supercooled regime where the relaxation times became  very long. On the other hand, hydrogen-bond connectivity and tetrahedrality agree to optimized four sites classical water models. Our results provide an interesting starting point to improve on the behavior of Drude based water models beyond ambient conditions.

\section{Methods}

\subsection{Simulation details}

All molecular dynamics simulations were run with the NAMD program  \cite{NAMD} with an integration step of 1~fs. The system contained 1024 water molecules in a cubic box. Temperature and pressure were controlled with a Langevin thermostat and Berendsen barostat with 1~ps and 100~fs relaxation time, respectively. The temperature of the Drude particles were set to 1~K at all conditions as suggested in the paper implementing the model into NAMD \cite{Jiang2010}. Non-covalent interactions were treated with a 1.2~nm cut-off and PME. Molecular trajectories of  50~ns in length were calculated for temperatures from 170~K to 260~K with steps of 10~K while  from 260~K to 340~K with steps of 20~K. At the higher temperatures (T$>$260~K)  the simulations length was of only 10~ns per trajectory because of the rapid equilibration times.

TIP4P/2005 simulations \cite{Abascal2005}  were run with the program GROMACS
\cite{gromacs} with an integration time-step of 2~fs. The water box
consisted of 1024  molecules in the NPT ensemble with pressure of
1 atm and temperatures ranging from 180~K to 350~K with steps of 10~K.
The Berendsen barostat \cite{Berendsen1984}, velocity rescale
thermostat \cite{Bussi2007} and PME \cite{Darden1993} were used for
pressure coupling, temperature coupling and long-range electrostatics,
respectively. 
The data was obtained from 1~ns long
simulations after 10~ns of equilibration in the NPT ensemble for T$>$240~K. For temperatures lower than 240~K 20~ns of equilibration was adopted.

\subsection{Hydrogen-bond propensities}

A maximum of four
hydrogen-bonds per molecule was considered with a bond being formed if the distance between oxygens and the angle O-O-H was smaller than 3.5~\AA\ and 30~degrees, respectively~\cite{Luzar1996}. Water structures were grouped into four archetypal configurations of population P$_i^{(*)}$ \cite{Shevchuk2012}: the fully coordinated first and second solvation shells for a total of 16 hydrogen-bonds (P$_4$); the fully coordinated first shell, in which one or more hydrogen bonds between the first and the second shells are missing or loops are formed (P$_4^*$); the three coordinated water molecule (P$_3$) and the rest (P$_{210}$). Within this representation the sum over the four populations is equal to one for each temperature.

\subsection{Tetrahedral order parameter}

The tetrahedral order parameter for a water molecule $i$ was calculated as

\begin{equation}
q_{i}=1-\frac{3}{8}\sum_{j=1}^{3}\sum_{k=j+1}^{4}\big(cos\psi_{jik}+\frac{1}{3}\big)^{2},
\end{equation}
where $j$ and $k$ are any of the four nearest water molecules of $i$ and $\psi_{jik}$ is the angle formed by their oxygens\cite{Errington2001}. The averaged value of the order parameter is denoted as Q.

\section{Results}

\subsection{The density maximum of the SWM4-NDP model}

\begin{figure}
  \includegraphics[width=80mm]{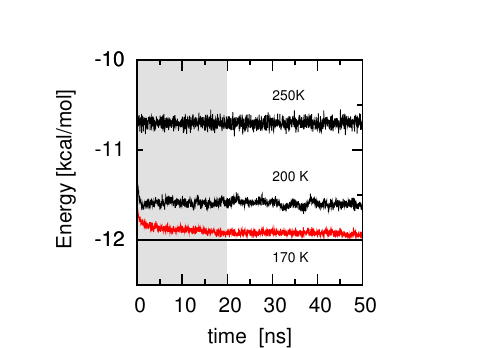}
  \caption{Timeseries of the potential energy of the SWM4-NDP water model for three different temperatures. Below 190~K relaxation times become very long as depicted by the 170~K trajectory (red)  and the gray region. }
  \label{fig:pot_en_ts}
\end{figure}

\begin{figure}
  \includegraphics[width=80mm]{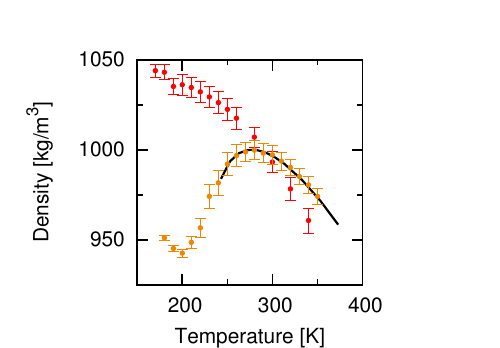}
  \caption{The density curve at 1 atm. Data for SWM4-NDP, TIP4P/2005 and experiments \cite{Kell1975} are shown in red, orange and black, respectively.}
  \label{fig:density_ts}
\end{figure}

Molecular dynamics simulations of the Drude-based polarizable water model SWM4-NDP were performed at several temperatures spanning from 170~K to 340~K. Running simulations for temperatures as low as 180~K, Kiss and Baranyai \cite{Kiss2012} recently showed that this model presents no density maximum. Independently from them we were also looking at the same problem. One important difference in our work is that simulations were run for much longer times: 50 ns per trajectory opposed to 5 ns in their case. 

Our results strongly indicate that long runs of several ns are needed to characterize  SWM4-NDP in the deeply supercooled regime. This becomes clear when looking at the time series of the potential energy. In Fig. 1A traces for different temperatures from 250~K to 170~K are shown. It was found that for temperatures lower than 200~K the relaxation time of the system dramatically slows down. The red line corresponding to  170~K shows that the system required at least 20~ns to equilibrate (gray region). This is a much longer time than the simulation length used in Ref.~\cite{Kiss2012}.

With the longer trajectories at hand, the density curve did present a maximum at around 200~K (red points in Fig.~2), a value that is similar to what was found for TIP3P (182~K \cite{Vega2005-2}). However, this maximum is not a global one as in experiments (black line) or in other classical models like TIP4P/2005 (orange points).  In fact, at lower temperatures (T$<$190~K) density grows again, making the density peak difficult to emerge from the statistical error, especially when using short trajectories. An increase of the density passed the density maximum is a feature of several water models. For example, this happens as well for TIP4P/2005 below 200~K (Fig.~2). What makes  SWM4-NDP peculiar is the fact that the value of the density in this regime becomes even higher than the density maximum, making the latter a relative maximum (not an absolute one).

\subsection{Hydrogen-bond propensities and temperature-shifts}

\begin{figure}
  \includegraphics[width=80mm]{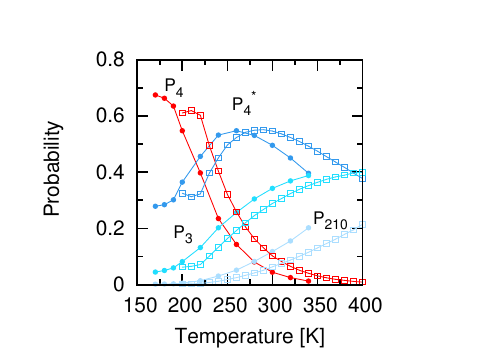}
  \caption{Hydrogen bond propensities including the second solvation shell. $P_4$, $P^*_4$, $P_3$ and $P_{210}$ are shown in red, dark blue, blue and light blue, respectively (see Methods for details). Data for SWM4-NDP and TIP4P/2005 are shown as filled circles and empty squares, respectively.}
  \label{fig:macro}
\end{figure}

Complementary information was obtained by investigating hydrogen-bond propensities. As done recently for seven classical water models \cite{Shevchuk2012} we calculated the probability to form fully coordinated hydrogen-bond configurations up to the second shell ($P_4$, red in Fig.~3; see Methods) as well as fully coordinated first shells with a disordered second shell ($P^*_4$, dark blue), three coordinated ($P_3$, blue) and less ($P_{210}$, light blue) first solvation shells. Results for the SWM4-NDP and TIP4P/2005 for comparison are shown in Fig.~3 as filled circles and empty squares, respectively. Contrary to the density analysis, hydrogen-bond propensities between the two models look much more similar (e.g. TIP3P showed a much more drastic temperature shift of 60~K \cite{Shevchuk2012}). The two sets of curves would nicely overlap if a shift of approximately 20~K is applied to the data. This observation suggests that while spatial rearrangement responsible for the density is dramatically different between the two models (and when compared to experiments), hydrogen-bond connectivity is similar. Such a discrepancy was already observed when comparing three-sites with four-sites models where a 10~K difference between temperature shifts estimated from hydrogen bonds or the position of the density maximum was observed \cite{Shevchuk2012}. But in this case the discrepancy is much larger being the temperature shifts respectively of 20~K and 80~K, i.e. a 40~K difference between the two approaches. 

\subsection{Tetrahedral order parameter}

\begin{figure}
  \includegraphics[width=80mm]{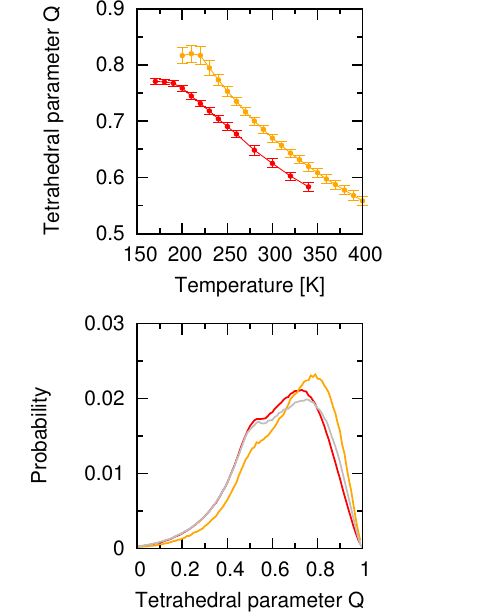}
  \caption{Tetrahedral order parameter. (Top) Average value of the parameter as a function of temperature. (Bottom) parameter distribution at 300~K. Data for SWM4-NDP and TIP4P/2005 are shown in red and orange, respectively.}
  \label{fig:tetra_temp}
\end{figure}

Temperature shifts were observed as well when comparing the two models on the base of the average value of the tetrahedral order parameter $Q$ (see Methods). The top panel of Fig.~4 shows this quantity as a function of temperature for both SWM4-NDP (red line) and TIP4P/2005 (orange). As for the case of the hydrogen-bond propensities $P_{i}^{(*)}$, the two models do not differ very much. At ambient conditions the temperature-shift is of about 30~K, a number that is in line with what observed for the hydrogen-bond propensities (Fig.~3). For the sake of comparison the distribution of $Q$ at 300~K for the two models is shown at the bottom of Fig.~4.  As it could have been expected from the behavior of the average value of the order parameter, TIP4P/2005 has a slightly larger fraction of molecules in a tetrahedral configuration but the overall shape of the distribution is similar for the two cases. This is even clearer when presenting the data for TIP4P/2005 at a 30~K higher temperature (gray curve): now the  distribution  for SWM4-NDP and the temperature-shifted TIP4P/2005 nicely overlap on top of each other with good approximation.  

\section{Discussion}

In the present work, we performed extensive molecular dynamics simulations of the Drude-based polarizable water model SWM4-NDP as a function of temperature. Contrary to what was reported in a recent paper \cite{Kiss2012}, it was found that the model do present a density maximum which was found to be around 200~K. The density curve was not easy to calculate because of the intrinsic slowing down of the system for temperatures lower than 200~K that hindered the detection of the maximum. To overcome this problem simulation runs of 50~ns each were performed, finding that at temperatures below 200~K the system required at least 20~ns to have the potential energy relaxing to a stationary average value without drifts. 

However, the density maximum we found is not as pronounced as other classical water models or in experiments. This was somewhat unexpected. As system temperature was lowered below 190~K, the density started to increase again. This is only in principle similar to what was observed for other models, like for example TIP4P/2005. In fact, in the present case the density value  increased to a value that is larger than the density maximum, making the latter a \emph{relative} maximum instead of an absolute one. The raising of the density at a such low temperature is probably due to some sort of frustration into the system leading to glassy behavior. This idea would also explain the dramatic slowing down of the relaxation kinetics of the model below 200~K.

In comparison to other classical models, SWM4-NDP performed very poorly in reproducing the density curve. This is somewhat disappointing given the success of other models in this respect, especially the reparametrized versions of the four-site model, TIP4P/2005 \cite{Abascal2005} and TIP4P-Ew \cite{Horn2004} as well as the newly presented iAMOEBA polarizable model \cite{Wang2013}.

Despite the position of the density maximum  of SWM4-NDP is shifted by roughly 80~K, the behavior of the hydrogen-bond propensities and tetrahedrality are very well in line to what the best models in the field predict. This behavior differs from what we found in the past for other non-polarizable water models, i.e. that a temperature-shift in the  density maximum corresponds  to a similar shift in  the hydrogen-bond propensities. The presence of polarizability instead completely decouples these two aspects, giving in principle a wider space to match experimental data, at least in principle.

In conclusion, our work shed some further light on the behavior of the SWM4-NDP polarizable model in temperature space. The great advantage of this model with respect to other approaches is the easy integration in all modern force-fields for biomolecular simulations. However, our results suggest that to make this model fully effective, a new parametrization able to  reproduce the density curve and other quantities in temperature  space is required.


\end{document}